\documentclass{article}
\usepackage{graphicx}

\title{Autoassociative Memory Retrieval and Spontaneous Activity Bumps in
Small-World Networks of Integrate-and-Fire Neurons}

\author{Anastasia Anishchenko\thanks{E-mail: anish@physics.brown.edu}\\
Department of Physics and Brain Science Program\\
Brown University, Providence RI 02912, USA
\and Elie Bienenstock\\
Division of Applied Mathematics, \\
Department of Neuroscience, Brain Science Program\\
Brown University, Providence RI 02912, USA
\and Alessandro Treves\\
Cognitive Neuroscience\\
SISSA - International School for Advanced Studies\\
Via Beirut 4, 34014 Trieste, Italy}

\begin{document}

\maketitle

\newpage
\begin{abstract}
The metric structure of synaptic connections is obviously an important
factor in shaping the properties of neural networks, in particular the
capacity to retrieve memories, with which are endowed autoassociative
nets operating via attractor dynamics.
Qualitatively, some real networks in the brain could be characterized as
'small worlds', in the sense that the structure of their connections is
intermediate between the extremes of an orderly geometric arrangement and of
a geometry-independent random mesh. Small worlds can be defined more
precisely in terms of their mean path length and clustering coefficient; but
is such a precise description useful to better understand how the type of
connectivity affects memory retrieval?

We have simulated an autoassociative memory network of
integrate-and-fire units, positioned on a ring, with the network
connectivity varied parametrically between ordered and random. We
find that the network retrieves when the connectivity is close to
random, and displays the characteristic behavior of ordered nets
(localized 'bumps' of activity) when the connectivity is close to
ordered. Recent analytical work shows that these two behaviours can
coexist in a network of simple threshold-linear units, leading to localized
retrieval states. We find that they tend to be mutually exclusive
behaviours, however, with our integrate-and-fire units. Moreover, the 
transition between the two occurs for values of the connectivity parameter 
which are not simply related to the notion of small worlds.
\end{abstract}

\newpage
\section{Introduction}

Autoassociative memory retrieval is often studied using neural network
models in which connectivity does not follow any geometrical pattern, -- i.e.
it is either all-to-all or, if sparse, randomly assigned. Such networks can be
characterized by their storage capacity, expressed as the maximum number of
patterns that can be retrieved, which turns out to be proportional to the number
$k$ of connections per unit. Networks with a regular geometrical rule informing 
their connectivity, instead, can often display geometrically localized patterns 
of activity, i.e. stabilize into activity profile 'bumps' of width proportional 
to $k$. Recently, applications in various fields have used the fact that 
small-world networks, characterized by a connectivity intermediate between 
regular and random, have different graph theoretic properties than either 
regular or random networks. In particular, small-worlds have both a high 
clustering coefficient $C$ and a short characteristic path length $L$, at the 
same time \cite{wattsstr}.

Note that while autoassociative retrieval is an obvious model for a
physiological memory process, also self-sustaining localized states
are of interest for the real brain, as models of the short-term memory
of a location on a cortical map. The two behaviours are combined in 
models of cortical function in which a localized and input-specific
pattern of activity represents the memory of a given stimulus in a given
position on the sensory array \cite{Tre03}.

Here we consider a family of 1D networks of integrate-and-fire
units, where changing the parameter of randomness $q$ allows to go
from a regular network ($q=0$) in which all connections are
relatively short-range, to a spatially random network ($q=1$) in
which all connections can be considered to be long-range.
Intermediate values of $q$ correspond to a biologically plausible
combination of long- and short-range connections. For $q=0$ such a
network spontaneously forms activity bumps \cite{sompo2}, and this
property is expected to be preserved at least for small values of
$q$. For $q=1$ the network behaves as an ordinary autoassociative
net \cite{amit,derrida}, and likewise this is expected to hold
also for $q<1$, as it does with a simpler binary-unit model
\cite{abramson}.

Can activity bumps and retrieval coexist?

We first explore how both the ability to retrieve, and to form bumps,
change with the degree of randomness $q$ in the connectivity.
Recent analytical work, on a model in which the connectivity has a Gaussian 
spread of variable width, in fact shows that retrieval and localization coexist
below a critical width value (and a critical storage load) \cite{Rou+04}.  
We can then address the question of whether the two abilities can coexist
also with our connectivity model,
at least in an intermediate regime between $q\simeq 0$ and $q\simeq 1$.

Is the transition between the different regimes related to small worlds?

Since the ability to form bumps depends on the mutual support among active units
through excitatory feedback loops, one can expect it to persist as long as
the clustering coefficient $C$ remains large, even as $q$ grows above 0.
Conversely, since pattern retrieval depends on the propagation of the retrieval
signal to all units in the network independently of their location, one
can expect robust retrieval to require that the mean path length from one unit 
to any other one is still short, even if $q$ is below 1. Thus, one
can expect an intermediate small-world regime of coexistence between
bumps and retrieval.

Can they coexist in more realistic networks, with integrate-and-fire dynamics?

Bohland and Minai \cite{minai}, and more recently McGraw and
Menziger \cite{mcgraw} and Morelli et al \cite{abramson} have
considered somewhat related issues with model networks of binary
neurons, in which localization and retrieval do not appear to coexist
(an observation explained by a simple argument proposed by Yasser Roudi
(personal communication)). By contrast, Roudi and Treves find that 
they coexist with threshold-linear units \cite{Rou+04}. The essential 
difference between these two very simple and dynamically implausible 
models, for the stability of bump states and of retrieval states, is 
whether single-unit activity levels saturate or not. Integrate-and-fire 
units with an absolute refractory period saturate to an intermediate 
and variable degree (depending, effectively, on how far below saturation 
they normally operate \cite{amit,treves}). They provide therefore a 
dynamically more plausible, but still very simple model to approach 
the issue of the coexistence of localization and retrieval  
in a cortical setting. This is what we discuss here, presenting
the result of computer simulations.

\section{Model}

\subsection{Integrate-and-Fire Model}

We consider a family of 1D networks of $N$ integrate-and-fire
units \cite{tuckwellbook}, arranged on a ring. In the simplest possible
integrate-and-fire model we adopt, the geometry of a neuron is
reduced to a point, and the time evolution of the membrane
potential $V_i$ of cell $i$, $i$ = 1$,\dots,N$ during an interspike
interval follows the equation
\begin{eqnarray}
\frac{dV_i(t)}{dt} &=& - \frac{V_i(t)- V^{res}}{\tau_m} +
\frac{R_m}{\tau_m} \left(I^{syn}(t) + I^{inh}(t) + I_i(t)\right)
\end{eqnarray}
where ${\tau}_m$ is the time constant of the cell membrane, $R_m$ its
passive resistance, and $V^{res}$ the resting potential. The total current
entering cell $i$ consists of a synaptic current $I^{syn}_i(t)$ due to the
spiking of all other cells in the network that connect to cell $i$, an
inhibitory current $I^{inh}(t)$, which depends on the network activity,
and a small external current $I_i(t)$, which represents all other inputs
into cell $i$.\\
\\
The last, external component of the current can be expressed through a
dimensionless variable $\lambda_i(t)$ as
\begin{equation}
I_i(t) = \lambda_i(t) (V^{th} - V^{res})/R_m
\end{equation}
\\
We use a simplified model for the inhibition, where it depends on the
excitatory activity of the network. The inhibitory current is the same for all
neurons at a given moment of time, and changes in time according to the
equation
\begin{eqnarray} {dI^{inh}(t)\over dt} &=& - \frac{1}{\tau_{inh}}
I^{inh}(t) +  \lambda^{inh}\frac{(V^{th} - V^{res})}{NR_m} r(t)
\end{eqnarray}
where $\lambda^{inh}$ is a dimensionless inhibitory parameter of
the model and $r(t)$ is a rate function characterizing the network
activity up to time $t$:
\begin{eqnarray}
r(t) &=& \sum_i \sum_{\{ t_i\} } \int_{-\infty}^t K(t-t')\delta
(t'- t_i) dt'
\end{eqnarray}
The kernel $K(t-t')$ we use for our simulations is a square wave
function lasting one time step $\Delta t$ of the integration, so that
$r(t)$ is just a total number of spikes in the time interval $\Delta t$. \\
\\
As $V_i$ reaches a threshold value $V^{thr}$, the cell emits a spike, after
which the membrane potential is reset to its resting value $V^{res}$ and held
there for the duration of an absolute refractory period $\tau_{ref}$.
During this time the cell cannot emit further spikes no matter how strongly
stimulated.

\subsection{Synaptic Inputs}

All synapses between neurons in the network are excitatory, and the synaptic
conductances are modeled as a difference of two exponents with time constants
${\tau}_1$ and ${\tau}_2$, such that ${\tau}_2$ is of the order of
${\tau}_m$ and ${\tau}_1>>{\tau}_2$. The synaptic current entering cell
$i$ at time $t$ then depends on the history of spiking of all cells that have
synapses on $i$:
\begin{eqnarray}
I^{syn}_i(t)  &=&  \frac{\tau_m}{R_m} \sum_{j{\neq}i} \sum_{\{ t_j\} }
\frac{\Delta V_{ji}}{\tau_1 - \tau_2}
\left( e^{-\frac{t - t_j}{\tau_1}} - e^{-\frac{t - t_j}{\tau_2}} \right)
\end{eqnarray}
where $\{ t_j \}$ are all the times before time $t$ when neuron $j$ produced
a spike and $\Delta V_{ji}$ is a size of the excitatory postsynaptic potential
(EPSP) in neuron $i$ evoked by one presynaptic spike of neuron $j$.\\
\\
The relative size of the evoked EPSP determines the efficacy of synapse
$ji$:
\begin{eqnarray}
\Delta V_{ji} &=&  \lambda^{syn}\frac{(V^{th} - V^{res})}{N}c_{ji} [1 + J_{ji}]
\label{efficacy}
\end{eqnarray}
where $\lambda^{syn}$ is the model parameter characterizing synaptic
excitation, $c_{ji}$ is $0$ or $1$ depending on the absence or presence of
the connection from neuron $j$ to neuron $i$, and $J_{ij}$ is modification of
synaptic strength due to Hebbian learning, which will be described later
(Section 2.4).\\

\subsection{Connectivity}

In our model, all connections in the network are uni-directional, so the
connectivity matrix is not symmetrical. We use a special procedure for
creating connections, after \cite{wattsstr}: the probability of one
neuron to make a connection to
another depends on a parameter $q$ in such a way, that changing $q$ from $0$ to
$1$ allows one to go from a regular to a spatially random network. \\
\\
A regular network ($q=0$) in this case is a network where connections are
formed based on a Gaussian distribution. The probability to make a
connection from neuron $i$ to neuron $j$ in such a metwork is
\begin{eqnarray}
P_0(c_{ij}) &=& e^{ -\frac{|i-j|^2}{2\sigma^2} }
\end{eqnarray}
where $|i-j|$ is the minimal distance between neurons $i$ and $j$ on the ring.
This results in an average number of connections per neuron $k \approx
\sqrt{2\pi}\sigma$ (actually, $k < \sqrt{2\pi}\sigma$ because of the finite
length of the ring).\\
\\
In a random network ($q=1$), the probability of creating a connection does
not depend on the distance between neurons:
\begin{eqnarray}
P_1(c_{ij}) &=& k/N
\end{eqnarray}
Intermediate values of $q$ thus correspond to a biologically plausible
combination of long- and short-range connections:
\begin{eqnarray}
P_q(c_{ij}) &=& (1-q)e^{ -\frac{|i-j|^2}{2\sigma^2} } + q \frac{k}{N}
\end{eqnarray}
The parameter $q$ is referred to as the `randomness' parameter,
following Watts and Strogatz, who used a similar procedure for
interpolating between regular and
random graphs. \\
\\
As originally analyzed by Watts and Strogatz\cite{wattsstr}, the
average path length and clustering coefficient in a graph wired in
this way both decrease with $q$, but their functional dependence
takes different shapes. Thus, there is a range of $q$ where the
clustering coefficient still remains almost as high as it is in a
regular graph, while the average path length already approaches
values as low as for random connectivity. Graphs that fall in this
range are called small-world graphs, referring to the fact that
for the first time this type of connectivity was explored in a
context of social networks \cite{milgram}. In this context the
average (or characteristic) path length $L$ can be thought of as
the average number of friends in the shortest chain connecting two
people, and the clustering coefficient $C$ as the average fraction
of one's friends who are also friends
of each other.\\
\\
In a neural network, $L$ corresponds to the average number of synapses that
need to be passed in order to transmit a signal from one neuron to
another, and thus depends on the number and distribution of long-range
connections. The clustering coefficient $C$ characterises the density of
local connections, and can be thought of as the probability of closing a trisynaptic loop
(the probability that neuron $k$ is connected to $i$, given that some neuron
$j$ which receives from $i$ connects to $k$).\\

\subsection{Hebbian plasticty}

Associative memory in the brain can be thought of as mediated by
Hebbian modifiable connections. Modifying connection strength
corresponds to modifying the synaptic efficacy of Eq.~\ref{efficacy}: \\
\begin{equation}
J_{ij} = \frac{1}{M} \sum_{\mu=1}^p
\left(\frac{\eta^{\mu}_i}{\langle\eta\rangle} - 1 \right)
\left(\frac{\eta^{\mu}_j}{\langle\eta\rangle} - 1 \right)
\end{equation}
where $\eta^{\mu}_i$ is the firing rate of neuron $i$ in the
$\mu$th memory pattern, $\langle\eta\rangle$ is the average firing
rate, the network stores $p$ patterns with equal strength, and $M$
is a normalization factor. The normalization factor should be
chosen to maximally utilize the useful range of synaptic efficacy,
which could be done by equalizing its variance with its mean. In
order to preserve the non-negativity of each efficacy, in Eq.~\ref{efficacy} we 
set $J_{ij}=-1$ whenever $(1+J_{ij})<0$.\\
\\
We draw patterns of sparseness $a$ from a binary distribution, so that
$\eta^{\mu}_i=1/a$ with probability $a$ and
$\eta^{\mu}_i=0$ with probability $1-a$. The 'average firing rate'
$\langle\eta\rangle$ in this case is equal to 1.\\

\subsection{Experimental Paradigm}

After the connections have been assigned and the synaptic efficacies modified by
storing memory patterns, we randomize the starting membrane potentials for all the
neurons:
\begin{equation}
V^{st}_i = V^{res} + \beta (V^{th} - V^{res})n_i
\end{equation}
where $\beta = 0.1$ is a constant and $n_i$ is a random number between 0 and
1, generated separately for each neuron $i$. We then let the network
run its dynamics.\\
\\
After a time $t_0$, a cue for one of the $p$ patterns (for instance,
pattern $\mu$) is given, in the form of an additional cue current injected
into each cell. The cue current is injected at a constant value up to the
time $t_1$, and then decreased linearly to 0 so that it becomes 0 at time
$t_2$. Thus, for the dimensionless current in (2) we have
\begin{equation}
\lambda_i(t) = \lambda^0 + \lambda^{cue}f(t) (\eta^{\mu}_i - 1)
\end{equation}
where $\lambda^0>0$ is a constant, dimensionless external current flowing
into all cells, $\lambda^{cue}$ is a dimensionless cue current parameter, and
\begin{equation}
f(t) = \left\{
\begin{array}
{c@{\quad: \quad}l}
0 & t<t_0 \quad $or$ \quad t_2 \le t\\
1 & t_0 \le t < t_1 \\
1-\frac{t - t_1}{t_2 - t_1} & t_1 \le t < t_2
\end{array}
\right.
\end{equation}
If the cue is partial -- for example, the cue quality $\rho = 25\%$ -- then
a continuous patch of the network, comprised of $25\%$ of the length along the
network ring, receives the cue for pattern $\mu$, whereas the rest of the
network receives a random pattern.
(In the simulations, the random pattern on the rest of the network has a
higher average activity than the original patterns, which makes the overall
network activity rise slightly during application of the cue.)\\
\\
During the time $t_{sim}$ of the simulation, spike counts
(numbers of spikes fired by each neuron), are collected in sequential time
windows of length $t_{win}$:
\begin{eqnarray}
r_i(t) &=&  \sum_{\{ t_i\}} \int_{t-t_{win}}^t \delta (t'- t_i) dt'
\end{eqnarray}
We then look at the geometrical shape of the
network activity profile in each window, and at the amount of overlap of this
activity with each of the patterns stored.\\
\\
The amount of overlap with pattern $\mu$ in the time window ending
at time $t$ is calculated as the cosine of the angle between the
pattern and the spike count vector:
\begin{eqnarray}
O^{\mu}(t) &=& \frac{ \sum_i \eta^{\mu}_i r_i(t) }
{ \sqrt{ \sum_i \left[ \eta^{\mu}_i \right]^2 \times
         \sum_i \left[ r_i(t) \right]^2 } }
\end{eqnarray}
\\
The measure of memory retrieval is
\begin{eqnarray}
m &=& \frac{1}{p}
      \sum_{\mu=1}^p
      \left[ \frac{O^{\mu}(t_{sim}) - O^{\mu}_{chance}(t_{sim})}
                                 {1 - O^{\mu}_{chance}(t_{sim})} \right]
\end{eqnarray}
where $O^{\mu}_{chance}(t)$ is the chance level of the overlap when the cue
for pattern $\mu$ is given:
\begin{eqnarray}
O^{\mu}_{chance}(t) &=& \frac{1}{p - 1} \sum_{\mu' \not= \mu} O^{\mu'}(t)
\end{eqnarray}
Note that the retrieval measure $m$ may occasionally become negative. This occurs when retrieval is so poor that, even though a cue for pattern $\mu$ has been given, the network activity happens to have less overlap with $\mu$ than it has, on average, with the other stored patterns.
\\
To quantify the prominence of the bump in the network activity profile, we
compute the standard deviation of the position of each neuron weighted by
its activity, with respect to the center of mass of the activity profile:
\begin{equation}
\sigma_a(t) = \sqrt { \frac {\sum_i |i-i_{CM}|^2 r_i(t)}{\sum_i r_i(t)} }
\end{equation}
The ratio of the standard deviation $\sigma_0$ of a uniform distribution to
$\sigma_a(t)$ has an intuitive meaning as the number of bumps that can be fit
on the ring next to each other. We take the measure of 'bumpiness' to be
\begin{eqnarray}
b(t) &=& \frac{ \frac{\sigma_0}{\sigma_a(t)} - 1 } { \frac{N}{k} - 1 }
\end{eqnarray}
For an easier comparison with the memory retrieval measure, we
compute the 'bumpiness' of a particular network in the last time window of
the simulation ($t = t_{sim}$), and then average over simulations with
different cues.
\\
\\
A rough measure of the relevance of saturation effects to the activity
of single units is the ratio $\psi$ between the firing rate of the most active
units in a localized state or in a retrieval state, and their maximum
possible firing rate, given by the inverse of the absolute refractory period.
In fact, for binary units $\psi = 1$ and for threshold-linear units $\psi=0$.
In our simulations, the most active units typically fired at 120 $Hz$ and
the absolute refractory period was 3 $msec$, implying $\psi \simeq 0.3$.
\\
\\
Table 1: Parameters Used for the Simulations.\\
\\
\begin{tabular}{|l|l|c|} \hline
Quantity & Symbol & Value \\[0.5ex]
\hline\hline
Number of cells & $N$ &  1000 \\
Average number of outgoing connections per cell & $k$ & 24 - 208 \\
Parameter of randomness & $q$ & 0 - 1 \\[0.5ex]
\hline
Synaptic excitation constant & $\lambda^{syn}$ & 40 \\
Inhibition constant & $\lambda^{inh}$ & 20 \\
External excitatory current parameter & $\lambda^0$ & 0.25 \\
Cue current parameter & $\lambda^{cue}$ & 0.1 \\[0.5ex]
\hline
Number of memory patterns stored & $p$ & 5 \\
Memory sparseness & $a$ & 0.2 \\
Synaptic plasticity normalization factor & $M$ & 10 \\
Cue quality & $\rho$ & 5 - 100\% \\[0.5ex]
\hline
Membrane time contant & $\tau_m$ & 5 $msec$ \\
Synaptic conductance time constant 1 & $\tau_1$ & 30 $msec$ \\
Synaptic conductance time constant 2 & $\tau_2$ & 4 $msec$ \\
Inhibition time constant & $\tau_{inh}$ & 4 $msec$ \\
Absolute refractory period & $\tau_{ref}$ & 3 $msec$ \\[0.5ex]
\hline
Integration time step & $\Delta t$ & 0.1 $msec$ \\
Simulation time & $t_{sim}$ & 1000 $msec$ \\
Cue onset time & $t_0$ & 150 $msec$ \\
Time when the cue starts to decrease & $t_1$ & 300 $msec$ \\
Cue offset time & $t_2$ & 500 $msec$ \\
Sampling time window & $t_{win}$ & 50 $msec$ \\[0.5ex]
\hline
\end{tabular} \\[1ex]
Note: Ranges are indicated for quantities that varied within runs.\\

\section{Results}

\subsection{Graph-Theoretic Properties of the Network}

\begin{figure}[h]
\caption{(A) Numerical results for the normalized characteristic path length
$L'(q) = [L(q) - L_{min}]/[L_{max} - L_{min}]$
and the normalized clustering coefficient
$C'(q) = [C(q) - C_{min}]/[C_{max} - C_{min}]$
as a function of the parameter of randomness (averaged over 10 network
realizations, all with the same number $k=41$ of outgoing connections
per neuron). Inset: logarithmic x-scale showing the small world regime for small $q$.
(B) Comparison of the analytical and numerical results for the clustering
coefficient $C$ (not normalized, $k=41$, $N=1000$).}
\label{Fig1}
\includegraphics[width=14cm]{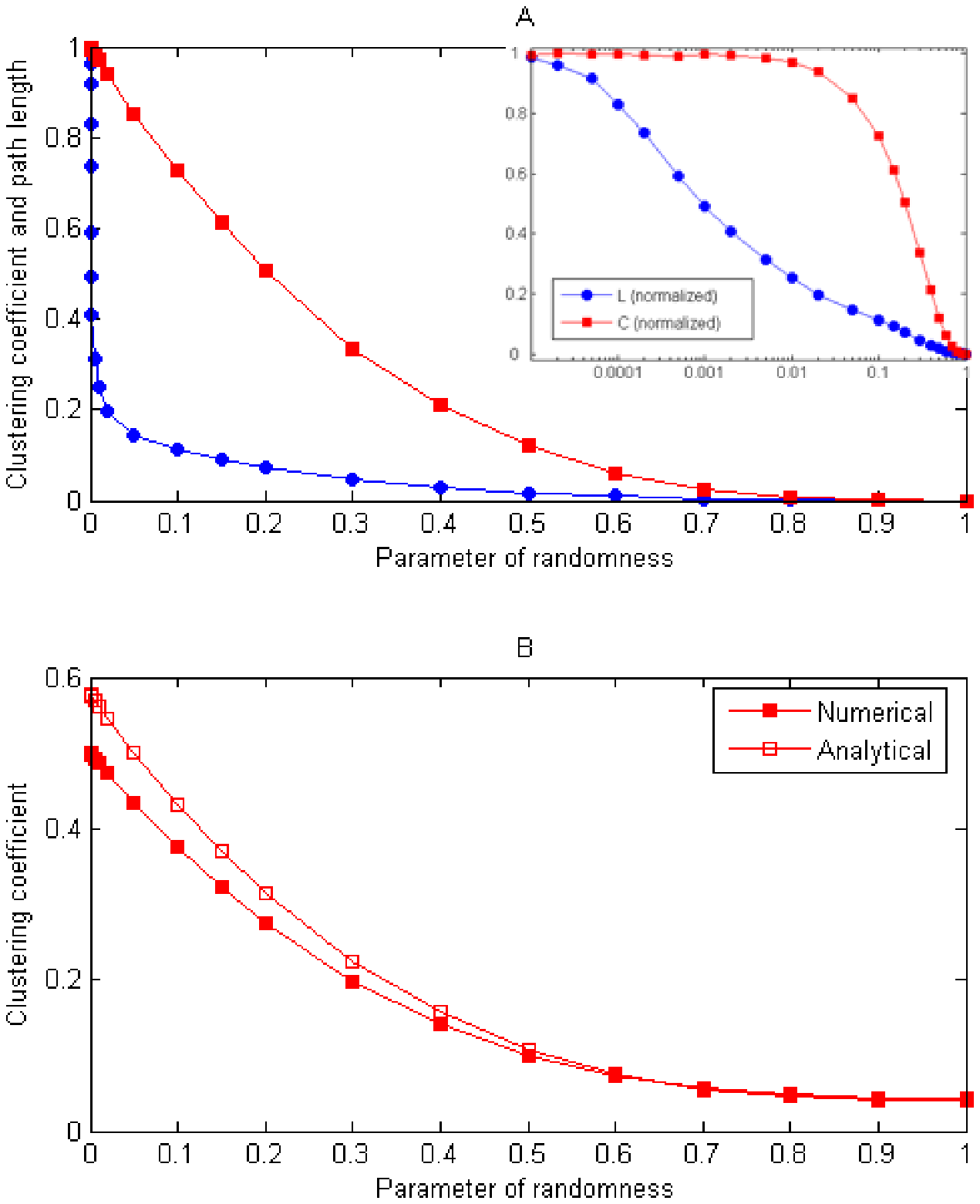}
\end{figure}

Finding an analytical approximation for the characteristic path length $L$ is
not a straightforward calculation. Estimating the clustering coefficient $C$ for a
given network is, however, quite easy. In our case, integrating connection
probabilities (assuming, for the sake of simplicity, an infinite network)
yields the analytical estimate
\begin{eqnarray}
C(q) &=& \left[ \frac{1}{\sqrt{3}} - \frac{k}{N} \right] (1 - q)^3 +
\frac{k}{N}
\label{canalyt}
\end{eqnarray}
or, after normalization,
\begin{eqnarray}
C'(q) = \frac{C(q) - C_{min}}{C_{max} - C_{min}} = (1-q)^3
\end{eqnarray}
which does not depend either on the network size $N$, or on the number $k$
of connections per neuron.\\
\\Once connections have been assigned, both $L$ and $C$ can be calculated numerically.
Keeping the number of connections per neuron $k > \ln N$ ensures that the graph
is connected \cite{bollobas}, i.e. that the number of synapses in the shortest
path between any two neurons does not diverge to infinity. \\
\\
The results of numerical calculations are shown in Fig.~\ref{Fig1}A.
The characteristic path length and the clustering coefficient both decrease
monotonically with $q$, taking their maximal values in the regular network
and minimal values in the random network. Similar to what was
described by Watts and Strogatz\cite{wattsstr}\cite{wattsbook}, $L$ and $C$
decrease in different fashions, with $L$ dropping down faster than $C$.
For example, for $q$ = 0.2,  $C'$ is at 52\% of its maximal value $C'(0)\equiv 1$,
while $L-L_{min}$ has already dropped to 8\% of $L(0)-L_{min}$. The inset shows
that $L$ approaches it maximal value $L(0)$ only for $q<10^{-4}$.\\
\\
Comparing the clustering coefficient estimated analytically with the simplified
formula (\ref{canalyt}) and the one calculated numerically shows some
minor discrepancy (Fig.~\ref{Fig1}B), which is due to the fact that
analytical calculations were performed for an infinite network. \\
\\
In conclusion, over a broad range $0.001\le q \le 0.2$ one can characterize
the mean path length as short, and the clustering coefficient as large, and one
can therefore expect simulations with e.g. $q=0.05$ or $q=0.1$ to demonstrate
characteristic small-world behaviour. \\

\subsection{Bumps Formation in a Regular Network}

\begin{figure}[h]
\caption{Spontaneous emergence of a bump in the activity profile of a
regular network ($q=0$) with an average number of outgoing connections per
neuron $k=41$.}
\label{Fig2}
\includegraphics[height=7cm,width=14cm]{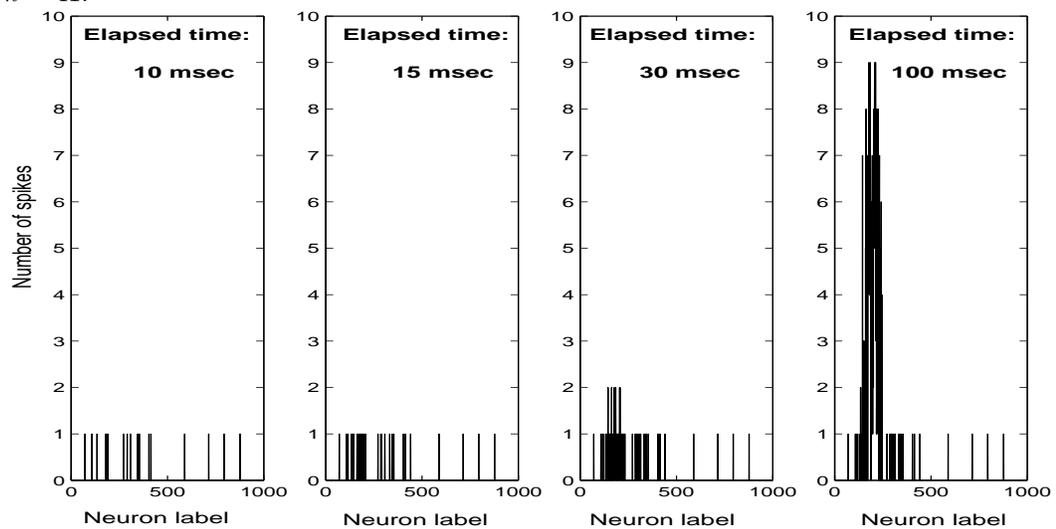}
\end{figure}

Networks whose connectivity follows a strict geometrical order
spontaneously form bumps in their activity profile, as shown in
Fig.~\ref{Fig2}. Bumps are formed before the presentation of a
pattern specific cue, and can either persist, or be temporarily or
permanently altered or diplaced after cue onset. Robust bumps,
which remain after the cue has been introduced and then removed,
tend to form only for regular or nearly regular connectivity
(small values of $q$). Fig.~\ref{Fig3}A exemplifies this behavior
for a network with $k=41$ connections per unit, $\rho = 0.25$ and different
values of $q$. In the example, a robust single bump forms for
$q=0$ and $q=0.2$. A double bump forms for $q=0.4$, and is
relocated to a different section of the ring following the cue. A
very noisy triple bump is barely noticeable towards the end of the
simulation for $q=0.6$, and no bumps can be really observed for
random or nearly random nets, $q=0.8$ and $q=1$. Thus bump
formation appears to be strongly favored by a regular
connectivity, as expected.\\

\subsection{Memory Retrieval in a Random Network}

\begin{figure}
\caption{ (A) Activity profiles during the simulation for different
values of the parameter of randomness. Shown is the number of
spikes $r_i$ (vertical axis) that each neuron (horizontal axis)
has fired in a sliding 50 $msec$ time window. (B) Overlaps
$O^{\mu}$ (vertical axis) of the network activity shown in (A)
with each of the $p=5$ previously stored memory patterns
(horizontal axis) \-- before, during, and after a 25\%-quality cue
for the pattern $\mu=2$ has been given.} \
\label{Fig3}
\includegraphics[width=12cm]{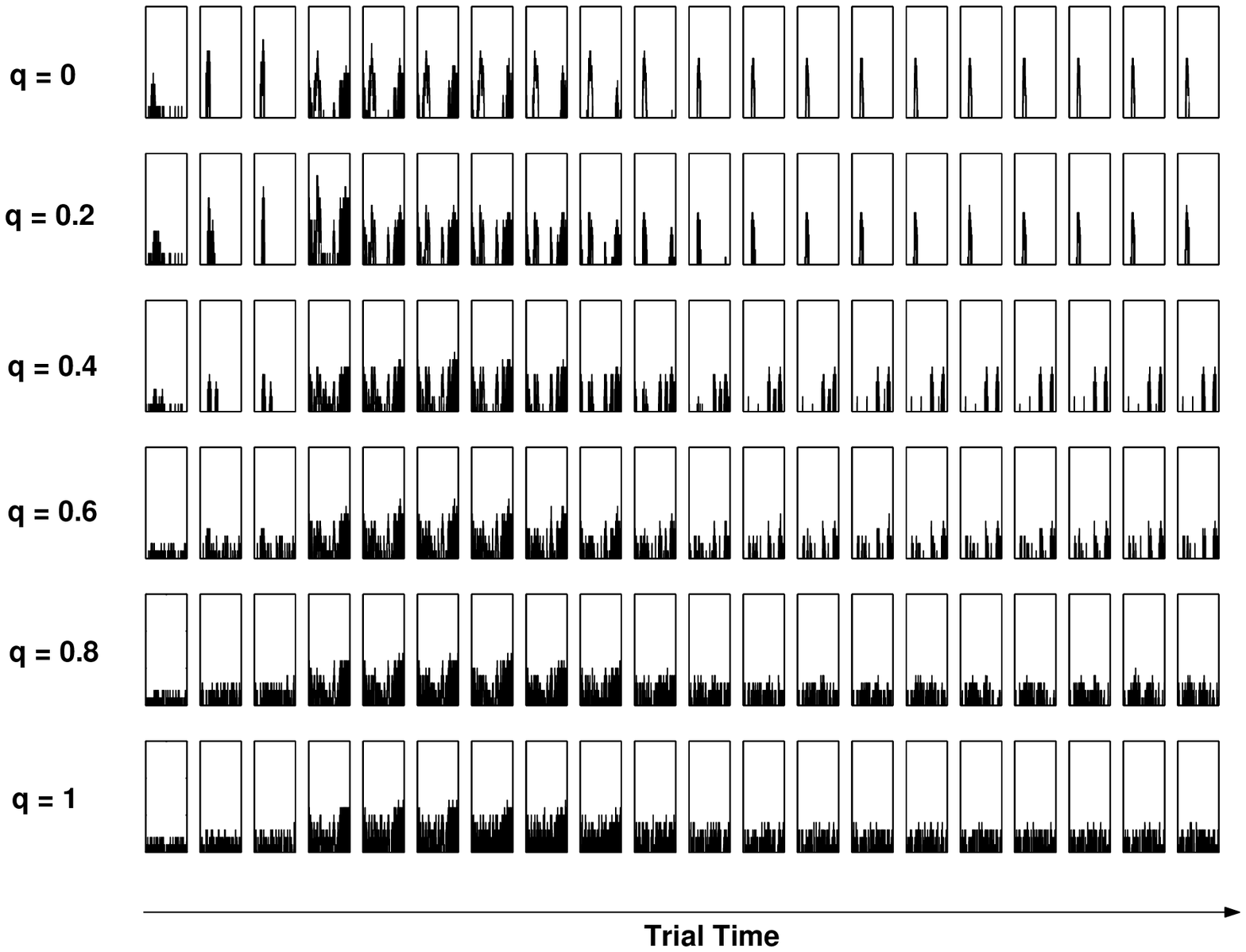}
\includegraphics[width=12cm]{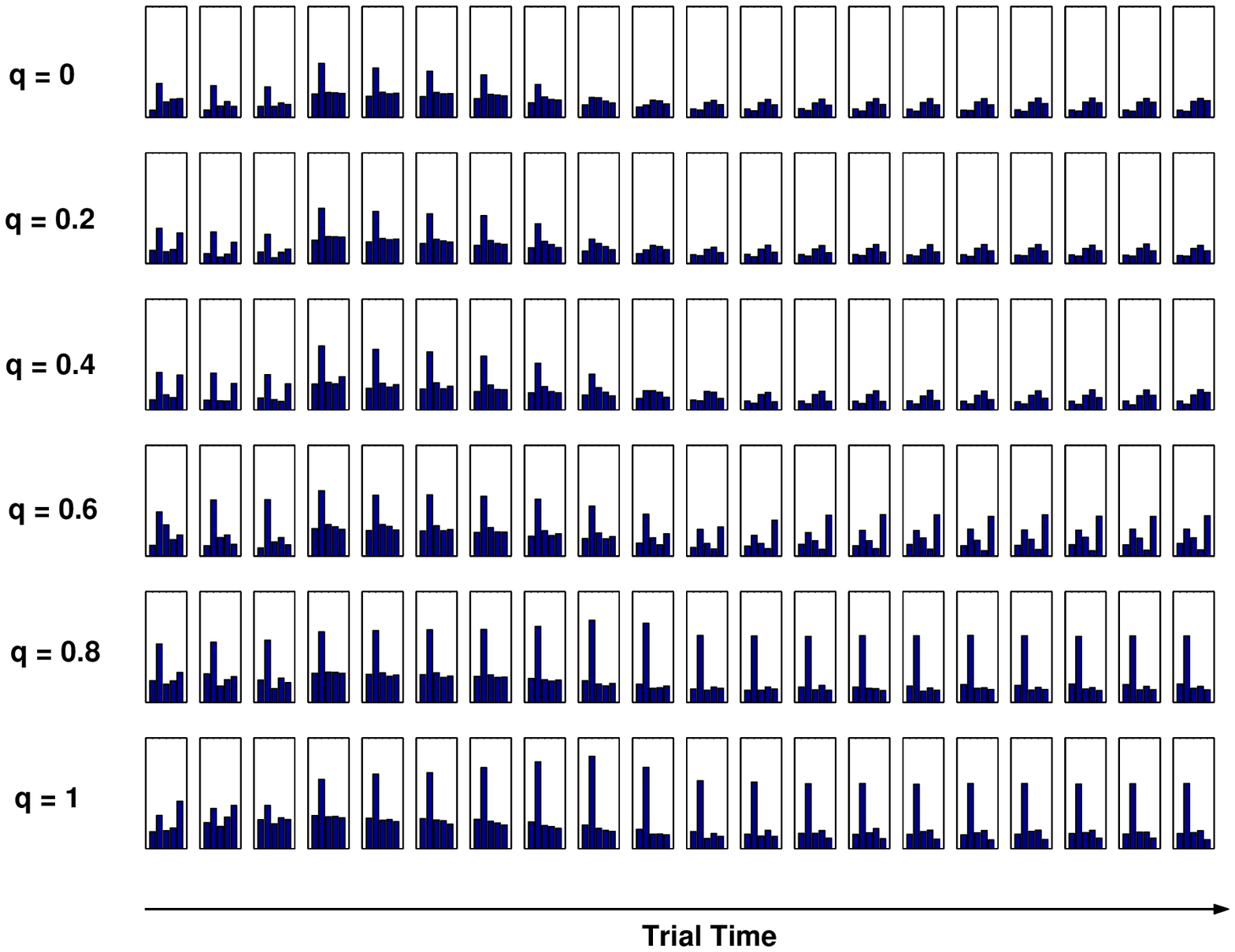}
\end{figure}

Fig.~\ref{Fig3}B shows the overlaps with the different memory
patterns in the same simulations used for Fig.~\ref{Fig3}A, with
$k=41$. The cue which is presented after $150msec$ is correlated
with the second pattern along one fourth of the ring, and it
reliably enhances the overlap with the second pattern during cue
presentation, whatever the value of $q$. By chance, the second
(and to some extent the fifth) pattern tends to have an elevated
initial overlap also before the presentation of the cue, a
behaviour often observed in autoassociative nets and due to
fluctuations. Despite this early 'advantage', once the pattern
specific cue has been completely removed (after $500msec$, that is
after the first 10 frames in the Figure) the overlap with the cued
pattern returns to chance level, except in the case of a fully
random ($q=1$) or nearly random ($q=0.8$) connectivity. Thus the
randomness in the connectivity appears to favor memory retrieval,
as expected, presumably by ensuring a more rapid propagation of
the retrieval signal to the whole network, along shorter mean
paths.

\subsection{Bumps and Memory in Networks of Different Degrees of Order}

\begin{figure}[h]
\caption{Bumpiness of the activity profile (A) and retrieval
performance (B) as a function of the parameter of randomness for
the average of 10 network realization, with $k = 41$ and $\rho=1$ (error bars are
report the stadard deviation across realizations).}
\label{Fig4}
\includegraphics[width=14cm]{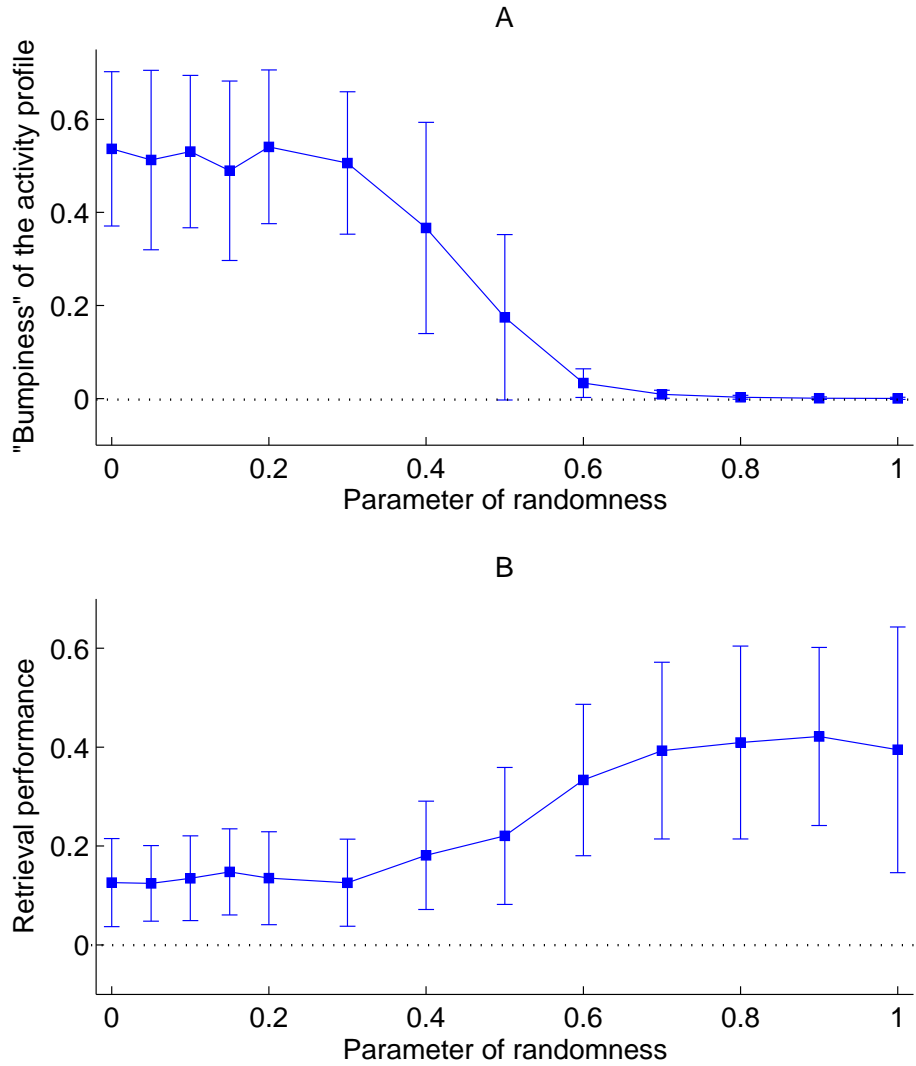}
\end{figure}

\begin{figure}[h]
\caption{The effect of the cue quality $\rho$ (displayed in the legend)
on retrieval performance, for the average of 10 network realizations with $k=41$. }
\label{Fig5}
\includegraphics[width=14cm]{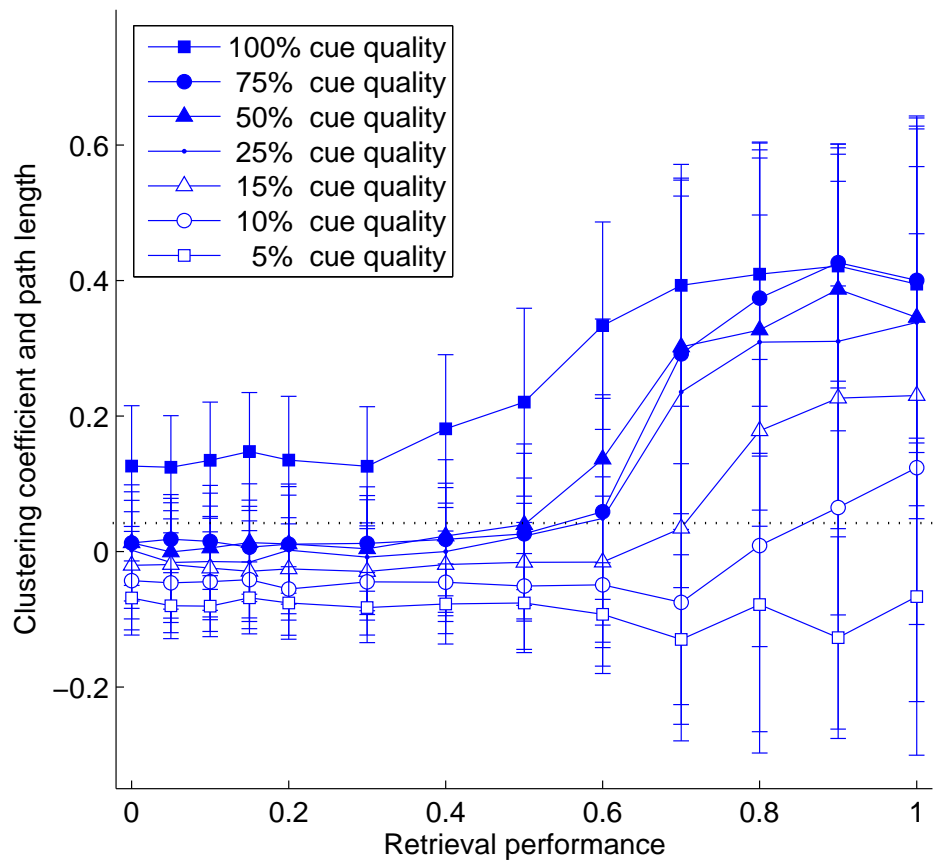}
\end{figure}

\begin{figure}[h]
\caption{The effect of changing the number $k$ of connections per
neuron on the retrieval performance (triangles) and on the
bumpiness (dots), for full cues.} 
\label{Fig6}
\includegraphics[width=16cm]{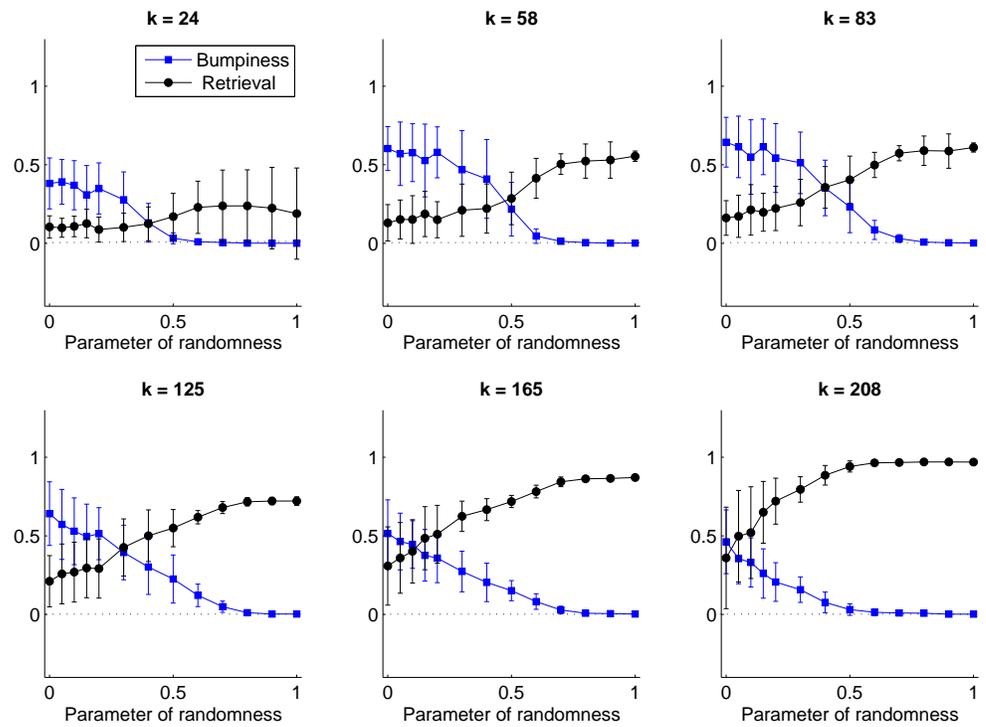}
\end{figure}

\begin{figure}[h]
\caption{The effect of changing the number $k$ of connections per
neuron on the retrieval performance (triangles) and on the
bumpiness (dots), for partial cues, $\rho=50\%$.} 
\label{Fig7}
\includegraphics[width=16cm]{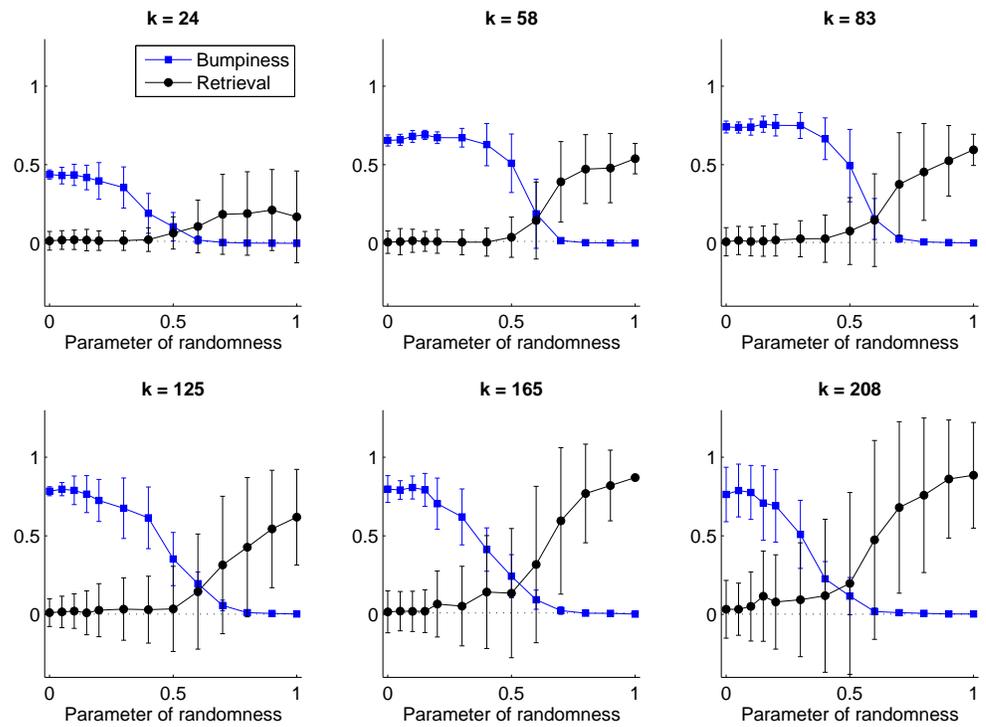}
\end{figure}

Fig.~\ref{Fig4} summarizes the results of simulations like those
of Fig.~\ref{Fig3}, for $k=41$ and a {\em fully extended} cue ($\rho =100\%$). 
The spontaneous activity bumps,
which are formed in the regular network, can be observed, somewhat
reduced, up to $q\approx 0.4\div 0.5$. For larger values of $q$ the bumpiness
measure quickly drops to a value close to zero. Storing random binary
patterns on the network does not affect the bumps, but the
retrieval performance appears to be very poor for small $q$, though the overlap
measure with the pattern to be retrieved remains marginally above chance.
As the randomness increases ($q \ge 0.6$), robust retrieval appears even though
the normalized overlap is far from its maximum at 1. Still, for this
value of the number of connections ($k=41$), and for full cues, the transitions 
from bumps to no bumps and from no retrieval to retrieval both appear to be
fairly sharp, and both occur between $q=0.4$ and $q=0.6$. \\
\\
Results are somewhat clearer when considering partial cues, as shown in Fig.~\ref{Fig5}, 
which reports the retrieval measure averaged over all 5 cued patterns and over 10
network realizations, and as a function of cue quality. Despite the considerable 
variability (see the large standard deviations), the average retrieval performance 
appears to depend smoothly on $q$ and $\rho$. For partial but still sufficiently large 
cues, $75\% \ge \rho \ge 25\%$, the network retrieves reasonably well when it 
is regular or nearly regular, $q\ge 0.7$, approaching the performance obtained 
with full cues. The overlap is reduced for $q=0.6$, and it approaches zero for $q\le 0.5$.
For $\rho=15\%$, retrieval appears for larger $q$ and is weaker, and for even lower 
cue quality, $\rho= 5\% - 10\%$, essentially no retrieval occurs. \\
\\
The non-zero average overlap measures reached by regular networks with full cues 
presumably reflect the network occasionally {\em freezing}, i.e. getting stuck in 
a particular firing configuration, what may be called remnant magnetization in a 
spin system. This makes the distinction between retrieval and no retrieval somewhat 
fuzzier. Nevertheless, whatever the size of the cue, it is apparent that
there is a well defined transition between two behaviors, bump
formation and retrieval, which for $k=41$ is concentrated between $q=0.5$ 
and $q=0.6$. The region of coexistence, if it exists, is in this case very limited. \\
\\
Changing $k$ does not affect the qualitative network behavior, as shown in Fig.~\ref{Fig6}
for the case of full cues and in Fig.~\ref{Fig7} in the case of partial cues ($\rho=50\%$).
Whatever the connectivity $k$ there is a transition between bump formation in nearly
regular networks (on the left in each panel) and retrieval in nearly random networks
(on the right). The transition is not always very sharp, and in some cases
one could describe an individual network as able to both form bumps and retrieve
memory patterns to some extent. When the number of connections is very limited ($k=24$),
both retrieval and bump formation are weak, when they at all occur. For more
numerous connections the degree of bumpiness, as quantified by our measure,
reaches for regular networks values close to 0.6 with full cues, and up to 0.75 with
partial cues. \\
\\
In Fig.~\ref{Fig6}, the range of the parameter of randomness, over which 
bump formation occurs, shrinks with increasing $k$, and in the case of $k=208$ 
the bumpiness for $q=0.2$ is already down to roughly 1/3 of its value for the 
regular network. The retrieval measure, instead, reaches higher values the larger
the connectivity, approaching 1 for $k=208$ and large $q$. In parallel to bump
formation, the range of $q$ values over which the network retrieves well appears 
to expand with $k$, so that for $k=208$ essentially only regular networks fail 
to retrieve properly. Overall the transition from no retrieval to retrieval 
appears to occur in the same $q$-region as the transition from bumps to no
bumps, and for the simulations with full cues reported in Fig.~\ref{Fig6} this 
$q$-region shifts leftward, to lower $q$ values, for increasing $k$.\\
\\
The same lack of a region of coexistence between bump formation and retrieval is 
evident in Fig.~\ref{Fig7}, reporting simulations with partial cues, $\rho=50\%$.
Again, the bumpiness measure and more prominently the retrieval measure
increase with more connections per unit, reflecting the larger storage capacity 
of networks with more connections. In our simulations the number of
patterns is always fixed at $p=5$, and the limited size of the network does not
allow a straightforward application of storage capacity calculations valid for large
systems \cite{Rou+04}. Still, networks with $k=24$ appear to operate above their
capacity, and also for larger $k$ the beneficial effect of a more extensive connectivity 
is in any case reflected in higher values for the overlap with the retrieved patterns. 
The major difference with Fig.~\ref{Fig6} is that with partial cues the location
of the transition between bumps and retrieval does not appear to be much affected 
by $k$, hovering between $q=0.45$ and $q=0.6$. Moreover, for regular and nearly
regular networks the normalized overlap does approach zero, without any remnant
magnetization effect; this appears to `allow' the bumpiness measure to approach
higher values, suggestive of an active interference between the two phenomena. 
Bump formation has indeed been analytically shown to lower the storage capacity of 
threshold-linear networks \cite{Rou+04}, and it makes sense that the converse be also
true, that is that even partial retrieval (which implies a proportion of quiescent units
in the would-be bump) should loosen the bump and weaken its localization.   \\
\\
These results of both Fig.~\ref{Fig6} and Fig.~\ref{Fig7} contradict the expectation 
that one or both of these transitions be simply related to the small-world property 
of the connectivity graph. As shown above, the normalized value of the clustering 
coefficient, whose decrease marks the right flank of the small-world regime, is essentially 
independent of $k$, as it is well approximated, in our model, by the simple function 
$(1-q)^3$. The transition between a localization regime and a retrieval regime, 
instead, to the extent that it can be defined for full cues (with the remnant
magnetization artefact) clearly depends on $k$; while for partial cues it occurs
at a degree of randomness clearly beyond the small world range. The transition 
does not, therefore, simply reflect a small-world related property.

\section{Discussion}

Although small-world networks have been recently considered in a
variety of studies, the relevance of a small-world type of
connectivity for associative memory networks has been touched upon
only in a few papers. Bohland and Minai \cite{minai} compare the
retrieval performance of regular, small-world and random networks
of an otherwise classic Hopfield model with binary units. They
conclude that small-world networks with sufficiently large $q$
approach the retrieval performance of random networks, but with the
advantage of a reduced total wire length. Note that wire length is
measured taking the physical distance between connected nodes into
account, while path length is just the number of nodes to be
hopped on in order to connect from one to another node of a pair.
This result of reduced wire length for similar performance is
likely quite general, but it relies on the small-world property of
the connectivity graph only in the straightforward sense of using
up less wiring (which is true also for values of $q$ above those
of the small-world regime); and not in the more non-trivial sense
of using also the high clustering coefficient of small-world
networks. The second result obtained by Bohland and Minai is that
small-world nets can 'correct' localized errors corresponding to
incomplete memory patters quicker than more regular networks (but
slower than more random network). This again illustrates behavior
intermediate between the extremes of a well performing associative
memory network with random connections, and of a poorly performing
associative memory network with regular connections.\\
\\
McGraw and Menziger \cite{mcgraw} similarly compare the final
overlap reached by associative networks of different connectivity,
when started from a full memory pattern (this they take to be a
measure of pattern stability) and from a very corrupted version of
a memory pattern (this being a measure of the size of a basin of
attraction). In terms of both measures small worlds again perform
intermediate between regular and random nets, and in fact more
similarly to random nets for values of $q\simeq 0.5$, intermediate
between 0 and 1. They go on to discuss the property of another
type of connectivity, the scale-free network, in which a subset of
nodes, the 'hubs', receive more connections than the average, and
therefore unsurprisingly performs better on measures of
associative retrieval restricted to the behavior of the hubs
themselves.\\
\\
In summary, neither of these studies indicates a genuine
non-trivial advantage of small-world connectivity for associative
memories.\\
\\
In our study, we considered integrate-and-fire units instead of
binary units, so as to make more direct contact with cortical networks. 
More crucially, we also investigated, besides associative memory, 
another behavior, bump formation, or localization, which can
also be regarded as a sort of emergent network computation.
Morelli et al in a recent paper \cite{abramson} study a seemingly analogous
phenomenon, i.e. the emergence, in an associative memory with
partially ordered connectivity, of discrete domains, in each of
which a different memory pattern is retrieved. Although broadly
consistent with our findings insofar as retrieval goes, the
phenomenon they study is almost an artefact of the binary units
used in their network, while we preferred to consider emergent
behavior relevant to real networks in the brain. We wondered
whether connectivity in the small-world regime might be reflected
in an advantage not in performing associative retrieval alone, but
in the coexistence of retrieval and bump formation. \\
\\
The possibility of such a coexistence is precisely the question that 
was addressed in a recent study that focused on analytically-tractable 
networks of threshold-linear units \cite{Rou+04}. While the connectivity
model was different (it had a Gaussian spread with varying width) and
unrelated to the notion of small-worlds, that study found that retrieval
states can indeed be localized, even in regular networks, at the price 
of a relatively minor decrease in storage capacity. The very feature that
makes threshold-linear units amenable to a full mathematical analysis,
the lack of saturation in their output, may however subtract from the
generality of the above results. In fact it was noted by Yasser Roudi (personal 
communication) that the units at the center of a {\em retrieval bump}
need to be activated to much higher levels than those at its flanks with
similar target activation in the memory pattern, or than the very same
units when they participate in a non-localized retrieval state. Localized
retrieval therefore requires the availability of a broad continuous range of 
activation levels (in the case of models with simple input-output units),
and it cannot occur with binary units, or even with sigmoidal units in
which active units all give essentially the same output.\\
\\
In our present model, we used biologically more relevant integrate-and-fire
units, whose firing rate, as a function of their inputs, follows, with
more complex dynamics, a roughly intermediate behaviour between binary
and threshold-linear. The absolute refractory period results in a saturation rate 
(in our case at 330$Hz$), but this ceiling is approached smoothly, and one has
to understand whether units can use the wide range of firing levels to 
accomodate retrieval specificity within spatially localized states.
We failed to observe this coexistence in practice, as networks go, with
increasing randomness parameter $q$, from a localization regime with no 
(or very poor) retrieval - to a retrieving regime with no bumps. Further, we
found that the critical transition $q$ values are beyond the boundaries 
of the small-world region.\\
\\
Because of its simulational character, our study cannot be taken
to be exhaustive. For example, the partial retrieval behavior
occurring for small $q$ might in fact be useful when sufficiently
few patterns are stored, $p$ small. In our simulations, $p=5$, but
then the whole network was limited in size and connectivity.
Moreover, a 2D network, which models cortical connectivity better
than our 1D ring, might behave differently. Further, memory
patterns defined over only parts of the network, which again
brings one closer to cortical systems, might lead to entirely
different conclusions about the coexistence of bump formation and
associative retrieval. Our simulations, in any case, indicate the importance of considering a 
realistic model of input-output neuronal transform while addressing
the issue of how to combine autoassociative memory retrieval with
localization on a cortical map, an issue of crucial importance to
understand the functions and microcircuitry of the cerebral cortex 
\cite{Tre03}.

\section*{Acknowledgments} We are grateful to Francesco P.
Battaglia and Haim Sompolinsky for their help at the early stage
of the project; to Yasser Roudi for extensive discussions of
his work; to the EU Advanced course in Computational
Neuroscience (Obidos, 2002) for the inspiration and opportunity to
develop the project;
and to the Burroughs Wellcome Fund for partial financial support.\\

\newpage

\newpage

\newpage

\newpage

\newpage

\newpage

%
%
%
%
%
%
%

\end{document}